\documentclass[runningheads]{llncs}

\usepackage[T1]{fontenc}
\usepackage[utf8]{inputenc}
\usepackage{amsmath,amssymb}
\usepackage{booktabs}
\usepackage{graphicx}
\usepackage{xcolor}
\usepackage{caption}
\usepackage{subcaption}
\usepackage{microtype}
\usepackage{enumitem}
\usepackage{bbding}
\usepackage{tabularx}
\usepackage{array}
\usepackage[numbers,sort&compress]{natbib}

\usepackage{hyperref}
\hypersetup{pdfauthor={Nils Bundi},pdfsubject={},pdfkeywords={},pdftitle={Pricing the DeFi Tail},hidelinks}

\graphicspath{{figures/}}

\newcommand{\E}{\mathbb{E}}

\newcommand{\VaR}{\operatorname{VaR}}

\newcommand{\NB}{\mathrm{NB}}

\begin{document}

\title{Pricing the DeFi Tail: Do Protocols or\\
Depositors Price Operational Risk?\thanks{Accepted at CBT 2026
(10th International Workshop on Cryptocurrencies and Blockchain
Technology), co-located with ESORICS 2026, and to be published by
Springer Nature in the ESORICS 2026 International Workshops
proceedings. This is not the Version of Record.}}
\titlerunning{Pricing the DeFi Tail}
\author{Nils Bundi\,(\Envelope)\orcidID{0000-0003-3576-3289}}
\authorrunning{N. Bundi}
\institute{Zurich University of Applied Sciences, School of Engineering, Winterthur, Switzerland\\
\email{bund@zhaw.ch}}
\maketitle

\begin{abstract}
Similar to banks, DeFi protocols expose depositors to operational risk
(USD~9.45~billion across 1{,}075 events since 2020). Unlike banks, they
are not required to hold capital against it. A protocol may maintain a
buffer voluntarily. Absent one, the risk falls on the depositor, who
should then demand a risk premium in the supply yield. I quantify the underlying tail on one
benchmark, a per-sector Basel loss-distribution approach fitted to a
new operational risk event dataset, and test both margins against it. Tails
in the four core sectors are no heavier than the Moscadelli banking band
$[0.85, 1.39]$. Bridge, Derivatives, and the residual Other sector exhibit
cyber-loss-level tails ($\hat\xi \approx 1.6$), with point estimates past
the infinite-mean boundary. The Lending tail implies a $\VaR_{99.9}$ capital buffer of
$18\%$ of TVL and of the ten largest Lending venues,
the four holding a buffer cover on average $5\%$ of it.
Under market discipline, depositors should demand a higher yield in compensation where
a venue does not maintain a buffer. I find that venues without a buffer pay a higher premium than those with (a $125$-bps gap in medians): evidence the market discriminates in the right direction. However, the premium falls far short of an adequately priced tail. This unpriced tail falls disproportionately on the retail
depositor, who sees only the posted rate but lacks the information and skills 
to price it. Because these products are not bank-regulated, I recommend 
disclosure over capital mandates: protocols, and any service
providers that front access to it, should publish standardized
losses, existing capital buffers and tail coverage.
\end{abstract}

\section{Introduction}

Decentralized finance (DeFi) offers yield-bearing \emph{earn products},
among them lending-pool supply positions, decentralized-exchange (DEX) liquidity-provision positions,
and yield vaults. Similar to a bank deposit, they expose the depositor
to \emph{operational risk}, the losses from failed processes, people,
systems, and external events. Unlike a bank, which must hold capital
against operational risk sized under Basel from a
\emph{loss-distribution approach} (LDA)
\citep{Moscadelli2004,Fontnouvelle2006,Cope2009,Chernobai2008}, a DeFi
protocol faces no such requirement, so no regulated buffer stands
between a loss event and the depositor's claim.

DeFi's architecture genuinely removes several bank-balance-sheet risks:
smart-contract custody replaces the custodian, settlement is atomic,
and counterparty credit risk collapses into the overcollateralization
embedded in the contract \citep{Werner2022,Auer2024}. But the
operational risks do not go away, and in some cases concentrate:
external attacks on contract code, insider-rugpull events, oracle and
economic-design manipulation, and configuration errors. Aramonte et
al.~\cite{Aramonte2021} call this the ``decentralisation illusion.''
The consolidated dataset assembled in this paper records
USD~9.45~billion of depositor-facing gross losses across 1{,}075
operational-risk events on DeFi protocols between 2020-02-11 and 2026-05-29.

The decisive difference is \emph{who bears that residual}. With no
deposit insurance and no resolution authority, in DeFi the depositor,
not the protocol, is on the hook. Faced with this heavy-tailed loss
process, the two participants can each respond, and this paper is
organized around their two responses.

\emph{The protocol can hold a capital buffer.} Some protocols maintain
governance-owned on-chain reserves with the operational-risk function
Basel calls capital: Aave's Umbrella safety module, the Sky surplus buffer, and Compound's per-market reserve factor. No
regulatory framework requires this; the buffers that exist are sized
by governance vote rather than by any loss model.

\emph{Where the protocol holds no buffer, the depositor should demand
a risk premium.} By the design of a lending market, depositors who
bear more operational risk should require a higher supply yield through
the borrow rate the market clears. Where that yield embeds a premium
over the risk-free rate sized to the operational-risk tail, depositors
bear the risk knowingly and are compensated for it: the same
market-discipline mechanism by which uninsured bank creditors price
institutional risk \citep{Flannery1998,Egan2017}. Market discipline
presumes the depositor can price the risk, which requires disclosure
the retail depositor typically lacks, a gap the policy discussion
returns to.

The two responses are substitutes for the same risk, made by
different participants, and neither is disciplined by a regulator.
This paper asks \emph{whether either response is sized adequately for
the operational-risk tail it is meant to cover}, and makes four 
contributions to answer it.

\emph{Contributions.} First, I assemble the most comprehensive
publicly-reproducible DeFi operational-risk event dataset I am aware of:
$n=1{,}075$ deduplicated events and USD~9.45~B of gross losses over
2020--2026, from seven public sources, filtered to DeFi
protocols and tagged by sub-sector and Basel Level-1 event type
\citep{BCBS2006}. Second, I quantify the loss process as a bank
supervisor would, fitting a per-sector LDA and benchmarking the tail
against banking operational risk
\citep{Moscadelli2004,Fontnouvelle2006,Cope2009}; the four core DeFi
sectors prove no heavier-tailed than banking, while Bridge, Derivatives, and
the residual Other sector exhibit cyber-loss-level tails
($\hat\xi \approx 1.6$; cf.~\cite{Eling2019}) with point estimates past the
infinite-mean boundary. Third, I test the
protocol's response (capital buffers) and the depositor's (risk premia)
against that common benchmark. Fourth, I find the two margins move
together in the right direction: venues holding a buffer pay a
significantly lower premium than those without (Mann--Whitney $p=0.01$),
evidence the market prices operational risk, but at an order of magnitude
below the modeled tail. I close with policy recommendations.

\emph{Organization.} Section~\ref{sec:lit} situates the
contribution within the literature, Section~\ref{sec:data} documents the
data and Basel taxonomy, and Section~\ref{sec:method} sets out the LDA
methodology. Section~\ref{sec:empirical} reports the empirical
per-sector severity, frequency, and capital fits, and
Section~\ref{sec:results} tests the protocol's and the depositor's
responses and whether they substitute. Section~\ref{sec:discussion} discusses the
substitution between them, implications, policy, and limitations, and
Section~\ref{sec:conclusion} concludes.

\section{Related Work}\label{sec:lit}

This paper sits at the intersection of three empirical literatures:
operational-risk modeling in banking, empirical DeFi security, and 
deposit pricing and market discipline.

\emph{Operational risk in banking.} The statistical study of
operational-risk losses begins with Moscadelli~\cite{Moscadelli2004},
who fit \emph{generalized Pareto distributions} (GPD) to the Basel
Committee's 2002 loss-data exercise and reported tail indices
$\hat\xi \in [0.85,1.39]$ across the eight business lines, establishing
heavy tails with infinite higher moments. de Fontnouvelle et
al.~\cite{Fontnouvelle2006} found tail indices near or above one in
internal-LDA models, with aggregate-loss \emph{value-at-risk} (VaR)
varying by an order of magnitude across severity choices. Cope et
al.~\cite{Cope2009} catalogued the resulting pitfalls and Chernobai et
al.~\cite{Chernobai2008} gave the canonical textbook treatment. Cyber
risk is the closest non-financial analogue: GPD cyber-loss fits reach
$\hat\xi \approx 1.60$, most subcategories in the infinite-mean regime
\citep{Eling2019}, with the tail itself evolving over time
\citep{Wheatley2016}. The Basel Committee ultimately replaced the
internal-model Advanced Measurement Approach (AMA) with the
standardized approach, citing model complexity and excessive capital
variability \citep{BCBS2017}. My DeFi findings reproduce that pathology.

\emph{Empirical DeFi security.} The DeFi-security literature is younger
but developing rapidly. Werner et al.~\cite{Werner2022} and Zhou et
al.~\cite{Zhou2023} provide systematization-of-knowledge surveys of the
attack surface; the Zhou corpus of 181 incidents to April 2022 is
closest in spirit to my dataset, though qualitative rather than
tail-fitting. Qin et al.~\cite{Qin2021} characterize flashloan-enabled
attacks and price-oracle manipulation, the mechanisms behind much of
the Lending and Stablecoin losses I measure; Perez and
Livshits~\cite{Perez2021} show that most statically vulnerable
contracts are never exploited, motivating the ex-post realized-loss
measurement I adopt. On the classification side, Weing\"artner et
al.~\cite{Weingartner2023} propose a ``risk wheel'' that sorts DeFi
risks into systematic and unsystematic classes, and Arora et
al.~\cite{Arora2026} build an FMEA risk-scoring framework---both
\emph{ex ante} qualitative frameworks, complementary to the \emph{ex
post} quantitative fit here. What this literature does not provide,
and what I add, is a deduplicated multi-source event dataset tagged to
the banking operational-risk taxonomy and fitted per sector.

\emph{Yield pricing and market discipline.} The yield leg draws on bank
market discipline. Flannery~\cite{Flannery1998} reviews evidence that
uninsured-creditor yields embed bank-condition information, and Egan et
al.~\cite{Egan2017} formalize the joint determination of deposit yields
and capital as substitutes. For DeFi, Heimbach et
al.~\cite{Heimbach2022} show Uniswap V3 LP fees often fail to compensate
for impermanent loss, and Cornelli et al.~\cite{Cornelli2025} document
that DeFi lending supply is driven by search-for-yield rather than
risk-based pricing. Both legs have been studied separately; the
substitution between protocol capital and depositor yield, which I
estimate directly, has not.

\emph{The gap.} Three gaps emerge. The DeFi-security literature
classifies risks and surveys incidents but publishes no deduplicated,
multi-source, event-level loss dataset tagged to the banking
operational-risk taxonomy; neither it nor the cyber-loss literature
reports per-sector severity and frequency parameters for DeFi; and no
work places bank market discipline and operational-risk capital side by
side as competing responses to a single quantified DeFi loss tail. This
paper closes all three.

\section{Data: A Consolidated DeFi Operational-Risk Dataset}\label{sec:data}

A credible LDA fit requires an extensive event-level loss dataset.
Previous DeFi-security and crypto-crime studies rely on a single
source (DefiLlama, SlowMist, Chainalysis, or a curated list), each
with known coverage gaps. I merge seven public feeds,
deduplicate across them, and tag every record against a DeFi sector
and the Basel Level-1 operational-risk taxonomy.

\emph{Sources.} The seven feeds are
\emph{(A)} the DefiLlama hack ledger \citep{Defillama};
\emph{(B)} the rekt.news leaderboard \citep{Rekt2026};
\emph{(C)} the SunWeb3Sec \emph{DeFiHackLabs} catalog
\citep{DeFiHackLabs2026};
\emph{(D)} the kismp123 \emph{DeFi-Security-Incident} corpus
\citep{Kismp2026};
\emph{(E)} the BlockSec Security Incidents Library \citep{BlockSec2026};
\emph{(F)} the de.fi/rekt-database \citep{DeFiRektDb2026}
($4{,}030$ records, 2011--2026); and \emph{(G)} the SlowMist Hacked
tracker \citep{SlowMistHacked2026} ($2{,}100$ records, 2012--2026).
Sources (F) and (G) are broadest and include out-of-scope memecoin
honeypots, CEX failures, and individual-user phishing, removed by the
filter below.

\emph{Consolidation.} Feeds are normalized to a common schema and
deduplicated in two passes: name clustering within a 21-day window,
then a same-date ($\pm 7$~d) and loss-within-$\pm 10\%$ pass that
catches cross-named duplicates (``Ronin Bridge''~/ ``Ronin Network'').
The reconciled loss is the \emph{median} of per-source amounts. A regex
removes out-of-scope records (CEX events, memecoins, NFTs, custodial-wallet failures,
Ponzi schemes, unattributed phishing, wallet-software exploits), and a second curated pass drops records that the regex missed through an explicit exclusion list. A total of 153 events is excluded as Non-DeFi this way.

\emph{Sector inference.} Each event is assigned to one of seven
sub-sectors: \emph{Bridge} (cross-chain message-passing and
token-bridging), \emph{Lending} (permissionless money markets),
\emph{DEX} (automated market makers and order-book exchanges),
\emph{Yield} (aggregators, vaults, and yield-tokenization),
\emph{Stablecoin} (algorithmic, collateralized, and basis-backed), 
\emph{Derivatives} (perpetual and option DEXs,
synthetic-asset issuers, and liquid-staking / restaking tokens), and \emph{Other} (residual DeFi protocols). The grouping follows the DeFi
financial-primitives typologies of Werner et al.~\cite{Werner2022}
and Auer et al.~\cite{Auer2024}. DefiLlama's category field
is used where present, else a regex chain resolves Bridge first, then
Lending, DEX, Stablecoin, Yield, and Derivatives. A curated per-sector
audit corrects residual mislabels (including DefiLlama tags that a
protocol's identity contradicts).

\emph{Risk-category inference.} Every record is tagged with a Basel
Level-1 event type, the canonical banking taxonomy defined in Basel II \citep{BCBS2006} 
and carried into the Basel III finalization. The five types that transfer to an
autonomous protocol are read as in Table~\ref{tab:basel}. Two additional 
categories, Employment Practices and Workplace Safety (EPWS) and Damage to Physical Assets (DPA), have no measurable on-chain analogue and are omitted. 
Textual evidence (technique, description, name) is applied first, with
DefiLlama's classification as fallback, so an operational signature
overrides a generic ``contract bug'' label.\footnote{In the February
2026 Moonwell cbETH event an oracle priced collateral at
USD~1.12 instead of $\approx$~USD~2{,}200, and liquidation bots
seized USD~1.78~m in bad debt within minutes. The source issue was an oracle 
misconfiguration (a team-side deployment error, EDPM), 
not a contract exploit (EF). See the Moonwell incident summary, 
\url{https://forum.moonwell.fi/t/mip-x43-cbeth-oracle-incident-summary/2068}.}

\begin{table}[tb]
\centering
\caption{Basel Level-1 operational risk event types and their interpretation in a DeFi protocol}
\label{tab:basel}
\small
\begin{tabular}{l p{3.4cm} p{7.4cm}}
\toprule
Code & Basel Level-1 & DeFi operational reading \\
\midrule
IF   & Internal Fraud & Rugpull, owner-drain, insider exit-scam, upgradeable-contract abuse, honeypot-as-protocol. \\
EF   & External Fraud & Smart-contract code bugs (reentrancy, access-control, rounding), credential compromise (key phishing), and infrastructure attacks (DNS hijack, approval phishing). \\
CPBP & Clients, Products, Business Practices & Economic-design and improper-practice attacks: flashloan-governance, oracle and price manipulation, donate-to-reserves, share-price inflation. \\
BDSF & Business Disruption, System Failures & Chain-level events: maximal extractable value (MEV), sequencer outage, consensus issue, chain halt. \\
EDPM & Execution, Delivery, Process Mgmt & Team-side configuration and deployment errors: oracle mis-deployments, broken governance parameters. \\
\bottomrule
\end{tabular}
\end{table}

\emph{Coverage and window.} The analysis window runs from
2020-01-01 \citep{Aramonte2021} to the 2026-05-29 data cutoff, giving the
$1{,}164$-event consolidated DeFi dataset of Table~\ref{tab:source-counts}
(the earliest in-sample event is dated 2020-02-11). A depositor-facing
filter then removes $89$ events (USD~$0.84$~B) whose losses are not borne by the
sector's earn user (e.g., governance-token mints, over-liquidation, treasury
drains), yielding the $1{,}075$-event working sample
(USD~$9.45$~B), of which $536$ ($50\%$) carry cross-source confirmation.

\begin{table}[tb]
\centering
\caption{Per-source contribution: raw record count and the number
entering the consolidated $1{,}164$-event DeFi dataset (after removing
non-DeFi records; deduplication yields fewer entries than the feeds' sum)}
\label{tab:source-counts}
\small
\begin{tabular}{l r r}
\toprule
Source & Raw records & In dataset \\
\midrule
DefiLlama hack ledger      &   528 & 379 \\
kismp123 DSI corpus        &   823 & 290 \\
BlockSec Incidents Library &   285 & 201 \\
rekt.news leaderboard      &   281 & 219 \\
DeFiHackLabs catalog       &   615 & 163 \\
de.fi/rekt-database        & 4{,}030 &   704 \\
SlowMist Hacked            & 2{,}100 &   639 \\
\midrule
Consolidated (deduplicated) & & 1{,}164 \\
\bottomrule
\end{tabular}
\end{table}

\section{Methodology: The Loss-Distribution Approach}\label{sec:method}

I quantify the DeFi operational-loss process per Sector with the standard
operational risk loss-distribution approach (LDA), the same tooling
a bank supervisor applies under Basel. The LDA has three parts: a
severity model for individual event losses, a frequency model for how
many events occur per year, and their compound aggregation into an
annual loss distribution. I follow Basel's original AMA and price the
tail with a $99.9\%$ VaR of that distribution \citep{BCBS2017},
adopted for comparability with the banking benchmark.

\emph{Severity.} I treat individual event gross losses as independent and identically distributed (iid) draws
from a severity distribution $F_X$ and model its upper tail by
\emph{peaks-over-threshold} (POT). For a threshold $u$ and excesses
$Y_i = X_i - u \mid X_i > u$, the GPD has density
\begin{equation}
g_{\xi,\beta}(y)
  \;=\; \frac{1}{\beta}\,\Bigl(1 + \xi\,\frac{y}{\beta}\Bigr)^{-(1/\xi + 1)},
  \qquad y>0,\ \beta>0,\ \xi\in\mathbb{R},
\end{equation}
fitted by maximum likelihood with multi-start Nelder--Mead. The
threshold quantile $q^*$ is chosen per sector by a plateau-stability
rule on the grid $\mathcal Q = \{0.50, 0.55, \dots, 0.90\}$ (the
highest $q$ at which $\hat\xi(q)$ is stable to within $\tau = 0.30$),
subject to two practitioner floors on the exceedance count $n_u$:
$n_u \ge 20$ (Embrechts et al.~\cite{EmbrechtsKM1997}) and
$n_u/n \ge 10\%$ (the DuMouchel~\cite{DuMouchel1983} ``top 10\%''
rule) and falling back to the lowest eligible quantile otherwise.

\emph{Frequency.} For each sector I fit a \emph{negative binomial} (NB) count
model $N_s \sim \NB(\mu_s, \alpha_s)$ to the monthly event series.
I chose the NB over a Poisson model because the
monthly event counts are strongly over-dispersed, and a per-sector
likelihood-ratio test rejects Poisson in five of the seven sectors
(Section~\ref{sec:empirical}). The NB is fitted at monthly
frequency. Annual counts entering the compound are then derived as the sum of twelve
i.i.d. monthly draws with dispersion
$\hat\alpha_{\text{annual}}=\hat\alpha_{\text{monthly}}/12$. Severity
and frequency are treated as independent, following standard banking-LDA
practice \citep{Chernobai2008,Cope2009}. A detailed analysis of this assumption and a potential joint model of clustered
severities is left to future work.

\emph{Compound aggregation.} The annual aggregate loss is
\begin{equation}
S_{\text{year}} \;=\; \sum_{i=1}^{N} X_i,
\qquad N \sim \NB(\hat\lambda_{\text{annual}}, \hat\alpha),
\qquad X_i \overset{\text{iid}}{\sim} F_X,
\end{equation}
where $\hat\lambda$ is the per-sector annual event rate, $\hat\alpha$
the NB dispersion, and $F_X$ the piecewise empirical-body / GPD-tail
mixture. Each simulated loss is capped at the largest single-protocol
exposure in the sector, since one operational event cannot lose more
than the assets at risk in one protocol. I draw $2\times 10^5$
simulated years per sector. The resulting mean $\E[S]$ is the
actuarial \emph{pure premium}, and $\VaR_{99.9}$ is the \emph{capital buffer} figure the Basel
AMA would compute. I express each sector's $\E[S]$ and $\VaR_{99.9}$
as a fraction of its trailing-365-day total value locked (TVL). TVL is the
natural exposure base for these events because an admin-key compromise, contract
exploit, or economic-design manipulation drains the smart contract's actual
holdings, not just the collateral pledged against active positions. In other words, every
dollar locked in the contract at the moment of the exploit is at risk. TVL
is thus the on-chain analogue to a bank's assets on books rather than its
risk-weighted assets. Two sectors lack a clean supply-side TVL, so I substitute
the closest exposure base: circulating supply for Stablecoin, and residual TVL (total
DeFi minus the five named sectors) for Other, both approximate and not strictly
comparable to the rest.

\emph{Per-protocol allocation.} No single protocol records enough
events for its own fit, and pooling losses across protocols of very
different size is a recognized challenge in operational-risk
measurement \citep{Cope2009}. Basel practice for banks lacking sufficient
internal-loss data prescribes the same fallback: a business-line-level
severity fitted collectively and allocated to the individual bank by an
exposure indicator \citep{Chernobai2008}. In the spirit of credibility
theory \citep{BuhlmannGisler2005}, I therefore fit once per sector and
treat its protocols as sharing that loss process per unit of TVL, a
collective estimate toward which data-poor venues are shrunk. Let
$\text{TVL}_p$ be the TVL in protocol $p$,
$\text{TVL}_s=\sum_{p\in s}\text{TVL}_p$ its sector total, and
$\lambda_s$ the sector event rate. The sector process then thins
to each protocol. A sector event strikes protocol $p$ with probability
$\text{TVL}_p/\text{TVL}_s$, so $p$ inherits a lower rate
$\lambda_p=(\text{TVL}_p/\text{TVL}_s)\,\lambda_s$ and the same
per-event severity, capped at its own exposure. Were severity uncapped the
mean would thin linearly and the pure premium would allocate by TVL share,
$\E[S]_p=(\text{TVL}_p/\text{TVL}_s)\,\E[S]_s$; the cap breaks that
invariance, so $\E[S]_p$ is simulated per venue and $\E[S]_p/\text{TVL}_p$
falls as a venue shrinks. The tail departs from linearity more sharply. With
$S_p$ the annual aggregate loss of protocol $p$, a bad year is
essentially one large event, and the single-loss approximation
$\VaR_{99.9}(S_p)\approx F_X^{-1}(1-0.001/\lambda_p)$
\citep{BockerKluppelberg2005,EmbrechtsKM1997} scales as
$\text{TVL}_p^{\hat\xi}$ with $\hat\xi<1$, so the VaR-to-TVL ratio rises
as a venue shrinks. I therefore simulate $\VaR_{99.9}$ per protocol
rather than scale the sector ratio, which understates the smaller
venues.

\section{Empirical LDA Fitting}\label{sec:empirical}

\begin{figure}[t]
\centering
\begin{subfigure}[t]{0.49\linewidth}
\centering
\includegraphics[width=\linewidth]{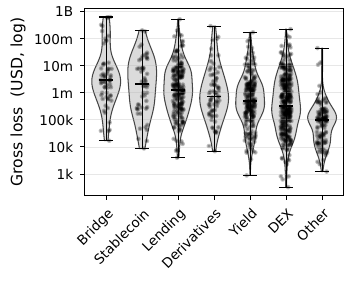}
\caption{Operational loss distribution; violins encode density, bars
the median.}
\label{fig:lossdist}
\end{subfigure}
\hfill
\begin{subfigure}[t]{0.49\linewidth}
\centering
\includegraphics[width=\linewidth]{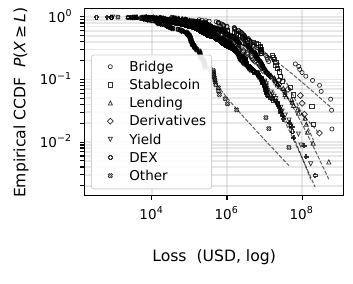}
\caption{Operational loss tails (Log-log CCDF), with fitted POT-GPD (dashed)}
\label{fig:ccdf}
\end{subfigure}
\caption{DeFi operational loss distribution and severity-tail fit across sectors}
\label{fig:severity}
\end{figure}

\emph{Data summary.} Losses are heavily right-skewed and differ sharply
across sectors, motivating a per-sector LDA. The aggregate mean is
USD~8.79~m against a median of USD~0.50~m and a maximum of USD~624~m
(Ronin). Sector totals span two orders of magnitude, from Bridge
(USD~3.24~B, $n=61$) to Other (USD~0.09~B, $n=121$). The per-sector
distributions in Figure~\ref{fig:lossdist} differ in location and shape,
Bridge carrying the highest median event loss (USD~2.90~m) from the
large pooled collateral cross-chain bridges hold. The dominant loss
mechanism differs too: sector and Basel event type are orthogonal, and
crossing them (Table~\ref{tab:sector-basel-usd}) shows Bridge and DEX as
near-pure EF stories (locked collateral value), Lending and Stablecoin as
CPBP (economic-design) failure heavy, and Derivatives as the only sector with significant BDSF exposure.

\begin{table}[tb]
\centering
\caption{Aggregate gross loss (USD~m) by sector $\times$ Basel
Level-1 type. The largest cell is Bridge$\times$EF (33\% of total).
Entries and margins are rounded independently.}
\label{tab:sector-basel-usd}
\small
\begin{tabular}{l r r r r r r}
\toprule
Sector $\backslash$ type & EF & IF & CPBP & EDPM & BDSF & Total \\
\midrule
Bridge       & 3{,}084 &  127 &     6 &   8 &  12 & 3{,}237 \\
Lending      &    704 &  568 &   928 &  14 &   8 & 2{,}222 \\
Stablecoin   &    178 &  105 &   456 &  27 &   8 &   773 \\
Yield        &    430 &  267 &   206 &  17 &   6 &   926 \\
DEX          & 1{,}044 &  136 &   221 &   0 &   1 & 1{,}402 \\
Derivatives  &    148 &    0 &   275 &   2 & 380 &   805 \\
Other        &     35 &    4 &    48 &   0 &   0 &    87 \\
\midrule
Total  & 5{,}623 & 1{,}207 & 2{,}140 &  68 & 415 & 9{,}452 \\
\bottomrule
\end{tabular}
\end{table}

\emph{Severity fits.} In Figure~\ref{fig:ccdf} the fitted GPD follows the
empirical CCDF across the exceedance range in every sector, excesses
scattering on both sides of the fit; in the $\hat\xi>1$ sectors it turns up
faster than the data at the extreme top, the divergence the exposure cap
addresses below. DeFi operational loss severity is thus heavy-tailed and the
banking literature's POT-GPD model applies. Table~\ref{tab:per-sector-pot} reports the per-sector
POT-GPD fit: Bridge ($\hat\xi=1.87$), Derivatives
($\hat\xi=1.45$), and Other ($\hat\xi=1.58$) fit above the infinite-mean
boundary. The four core sectors (Lending,
Stablecoin, Yield, DEX) cluster tightly in $\hat\xi \in [0.61, 0.75]$,
no heavier than the Moscadelli (2004) banking band $[0.85, 1.39]$ and
all finite-mean. With small exceedance samples in every sector ($20 \leq n_u \leq 51$),
these point estimates have to be interpreted with caution, as the wide
confidence intervals show. I thus ask whether the lognormal model---an alternative severity model of
the banking operational-risk literature
\citep{Fontnouvelle2006,Chernobai2008}---would offer a better fit, and report the Vuong \citep{Vuong1989} non-nested test (evaluated at the POT threshold $u^*$) in Table~\ref{tab:per-sector-pot}.
The test statistic returns a tie in every sector: $|V| \le 1.37$ against a $5\%$ critical value of $1.96$, with $p \ge 0.17$ throughout. The lognormal therefore does not fit these exceedances better than the GPD does, and at these exceedance counts the data do not separate the two families at all. I therefore retain the GPD model.

\begin{table}[tb]
\centering
\caption{Per-sector POT-GPD fits with $1{,}000$-replicate bootstrap CIs
on $\hat\xi$ at the plateau-selected quantile $q^*$ ($n_u$ exceedances);
the last two columns give the Vuong non-nested test of
the GPD against a lognormal, both fitted by
maximum likelihood to the same exceedances ($V>0$ favors the GPD;
$|V|>1.96$ separates them at $5\%$).}
\label{tab:per-sector-pot}
\footnotesize
\setlength{\tabcolsep}{4pt}
\begin{tabular}{lrrrlrrr}
\toprule
Sector  & $n$ & $q^*$ & $n_u$ & $\hat\xi$ \,[95\% CI] & $\hat\beta$ (USD m) & $V$ & $p$ \\
\midrule
Bridge        &      61 & 0.65 & 20 & +1.87 \,[$+0.58$, $+3.18$] &  17.6 & $-1.37$ & 0.17 \\
Lending       &     207 & 0.90 & 21 & +0.75 \,[$-0.28$, $+1.50$] &  24.8 & $-1.32$ & 0.19 \\
Stablecoin    &      52 & 0.60 & 21 & +0.65 \,[$-0.39$, $+1.33$] &  13.2 & $-0.03$ & 0.97 \\
Derivatives   &      71 & 0.65 & 25 & +1.45 \,[$+0.42$, $+2.38$] &   3.7 & $-1.04$ & 0.30 \\
Yield         &     222 & 0.90 & 23 & +0.61 \,[$-0.22$, $+1.24$] &  11.5 & $-0.74$ & 0.46 \\
DEX           &     341 & 0.85 & 51 & +0.72 \,[$+0.18$, $+1.15$] &   7.1 & $-0.02$ & 0.98 \\
Other         &     121 & 0.80 & 24 & +1.58 \,[$+0.54$, $+2.63$] &   0.2 & $+0.57$ & 0.57 \\
\bottomrule
\end{tabular}
\end{table}

\emph{Frequency fits.} The monthly counts are strongly over-dispersed (cross-sector dispersion index $D = 3.27$)---events cluster rather than arrive independently---so a Poisson model would understate the annual aggregate and its $\VaR_{99.9}$. Instead, I use the negative binomial (NB) model; per-sector fits are reported in Table~\ref{tab:nb} and give annual event rates from $8.3$/yr (Stablecoin) to $54.2$/yr (DEX). The table further reports results of a per-sector likelihood-ratio (LR) test that rejects Poisson at $p \le 0.02$ in five of seven sectors; it does not reject
for Stablecoin ($p = 0.59$), the sector with the fewest events, and for
Derivatives ($p = 0.05$) rejects it almost at $p \le 0.05$, confirming NB is overall the better choice.

\begin{table}[tb]
\centering
\caption{Per-sector negative-binomial frequency fits. $n$ is the sector's
event count; $\hat\mu_s$ the fitted mean monthly count and
$\hat\lambda_s = 12\hat\mu_s$ the implied annual event rate entering the
compound; $\hat\alpha_s$ the monthly NB dispersion. LR and $p$ give the likelihood-ratio statistic against a Poisson and its $p$-value
($\chi^2_1$).}
\label{tab:nb}
\small
\begin{tabular}{l r r r r r r}
\toprule
Sector & $n$ & $\hat\mu_s$ & $\hat\lambda_s$ & $\hat\alpha_s$ & LR & $p$ \\
\midrule
Bridge       &  61 & $0.80$ &  $9.7$ & $0.72$ & $6.2$   & $0.013$ \\
Lending      & 207 & $2.72$ & $32.9$ & $0.26$ & $10.2$  & $0.0014$ \\
Stablecoin   &  52 & $0.68$ &  $8.3$ & $0.14$ & $0.3$   & $0.590$ \\
Derivatives  &  71 & $0.93$ & $11.3$ & $0.41$ & $3.7$   & $0.054$ \\
Yield        & 222 & $2.92$ & $35.3$ & $0.60$ & $42.3$  & $7.7\times10^{-11}$ \\
DEX          & 341 & $4.49$ & $54.2$ & $0.31$ & $38.5$  & $5.4\times10^{-10}$ \\
Other        & 121 & $1.59$ & $19.2$ & $0.89$ & $30.4$  & $3.6\times10^{-8}$ \\
\bottomrule
\end{tabular}
\end{table}

\emph{Per-sector capital.} Combining severity and frequency, the
compound LDA gives the per-sector capital buffer $\VaR_{99.9}$ of
Table~\ref{tab:lda-sector}. Each simulated single-event loss is capped
at the largest single-protocol exposure in its sector, since one
operational event cannot drain more than the assets at risk in one
protocol. The core sectors Lending, DEX, Yield, and Stablecoin produce $\VaR_{99.9}$ ratios of
$14$--$18\%$ of their exposure bases, well above the operational-risk
capital banks hold \citep{Fontnouvelle2006}. Bridge, Derivatives and Other carry the largest requirements ($23$--$32\%$ of TVL), but only because each event is capped at exposure. Their
$\hat\xi>1$ point estimates put severity in the infinite-mean regime,
where no stable quantile exists, so uncapped the Bridge $\VaR_{99.9}$
diverges to $673\times$ TVL (Table~\ref{tab:lda-sector},
$\VaR_{99.9}^U$). The heaviest-tailed, most consequential sectors are
thus exactly the ones the model cannot pin down unaided, the instability
that retired the Basel internal-model AMA in 2017 \citep{BCBS2017}.

\begin{table}[tb]
\centering
\caption{Per-sector LDA capital ($\VaR_{99.9}$, \% of trailing-365d TVL),
compounded from the severity and frequency fits. $\VaR_{99.9}^U$ is the same simulation with the exposure cap removed, and IQR
the parametric-bootstrap interquartile range (200 replicates from the fitted
body+GPD mixture, refit and simulated at the same $q^*$).
$^\dagger$circulating supply, $^\ddagger$residual TVL.}
\label{tab:lda-sector}
\small
\begin{tabular}{l r r r r l}
\toprule
Sector & $n$ & TVL (B) & $\VaR_{99.9}$ (\%) & $\VaR_{99.9}^U$ (\%) & IQR (\%) \\
\midrule
Bridge       &  61 &  49.1  & 32.5 & 67{,}326 & [25.5,\,40.8] \\
Lending      & 207 &  66.8  & 18.0 & 22.3 & [3.2,\,18.9] \\
Stablecoin   &  52 & 26.5$^\dagger$ & 16.1 & 16.1 & [3.1,\,30.1] \\
Derivatives  &  71 &  65.9  & 22.8 & 638.0 & [21.8,\,27.1] \\
Yield        & 222 &  11.4  & 15.3 & 29.5 & [5.5,\,17.3] \\
DEX          & 341 &  14.5  & 14.1 & 46.2 & [10.5,\,15.9] \\
Other        & 121 &  18.2$^\ddagger$ & 27.5 & 285.9 & [4.8,\,27.7] \\
\bottomrule
\end{tabular}
\end{table}

\emph{Robustness.} Tail-index estimates are known to be sensitive to a handful of extreme
observations \citep{EmbrechtsKM1997}. This is confirmed in my sample: dropping the single largest event
moves $\hat\xi$ by $0.16$--$0.5$ across sectors, most sharply in Lending
($\hat\xi=0.75 \to 0.35$). Under any drop-top rule up to $0.5\%$, however,
every sector retains its below/above infinite-mean characteristic. A cap-sensitivity check
(Table~\ref{tab:lda-sector}, $\VaR_{99.9}^U$ column) shows that on the
net basis the exposure cap binds for the reported sectors. Removing it
raises Lending from $18\%$ to $22\%$, DEX from $14\%$ to $46\%$, and
Yield from $15\%$ to $30\%$ of TVL. Their $\VaR_{99.9}$ should therefore
be read as cap-bounded, the $99.9\%$ year being dominated by the
largest venue's at-risk funds failing. Only Stablecoin is cap-insensitive, its $99.9\%$ loss
falling below the single-protocol cap. Bridge
and Derivatives ($\hat\xi>1$) are stable only under the cap. Without it
the Bridge figure diverges to $673\times$ TVL and Derivatives rises to
$638\%$. The conclusions are also robust to the
confidence level. Lowering the benchmark to $99.5\%$ and $99\%$ cuts the
Lending sector capital from $18\%$ to $7\%$ and $5\%$ of TVL and lifts
buffered-venue mean coverage from $5\%$ to $18\%$ and $31\%$, so even at
$99\%$ the buffer covers only about a third of a requirement still an
order of magnitude above bank operational-risk capital \citep{Fontnouvelle2006}.

\section{Bearing the Tail: Capital Buffers and Depositor Premia}\label{sec:results}

An earn product's operational-risk tail must be borne by someone.
The protocol can fund it ex-ante by holding a capital buffer, or the
depositor can be paid to bear it through a risk premium in the supply
yield. I measure each margin against the per-sector $\VaR_{99.9}$
capital buffer benchmark of Section~\ref{sec:empirical}, then ask
whether the two substitute.

\emph{The protocol's response.} The first response is to hold capital
against the tail. Applying the per-protocol TVL-share allocation of
Section~\ref{sec:method} to the ten largest Lending venues by gross supplied TVL
(DefiLlama \citep{Defillama}, June~2026), four of the ten disclose an
operational-risk reserve and six do not
(Table~\ref{tab:protocol-adequacy}). Under the per-sector homogeneity
assumption, the four buffered venues cover on average only $\sim 5\%$ of
their modeled capital buffer (a $95\%$ average shortfall), ranging from
$2\%$ (Venus) to $9\%$ (Aave~V3). The inadequacy is specific
to the tail only though, all four buffers cover the pure premium (expected annual
loss) at least once over, from Venus ($187\%$) to Compound~V3 ($700\%$)
(Table~\ref{tab:protocol-adequacy}).

\begin{table}[tb]
\centering
\caption{The four top-10 Lending venues holding a buffer (top-10
selected by gross supplied TVL, June~2026; TVL column is net, the funds a
contract actually holds). Parentheses give buffer coverage of per-protocol $\E[S]$
and $\VaR_{99.9}$. Buffer data sources: $^a$\texttt{aave.com};
$^b$\texttt{skyeco.com}; $^c$\texttt{compound.woof.software};
$^d$\texttt{bscscan.com}.}
\label{tab:protocol-adequacy}
\small
\begin{tabular}{l r r r r}
\toprule
Protocol & TVL (B) & Buffer (B) & $\E[S]$ (B) & $\VaR_{99.9}$ (B) \\
\midrule
Aave V3         & 11.85 & 0.36\,$^a$ & 0.08 ($475\%$) & 4.00 ($9\%$) \\
SparkLend       &  3.41 & 0.08\,$^b$ & 0.02 ($390\%$) & 1.58 ($5\%$) \\
Compound V3     &  1.04 & 0.04\,$^c$ & 0.01 ($700\%$) & 0.64 ($6\%$) \\
Venus Core Pool &  0.98 & 0.01\,$^d$ & 0.01 ($187\%$) & 0.61 ($2\%$) \\
\bottomrule
\end{tabular}
\end{table}

\emph{The depositor's response.} Where the protocol holds no buffer, or
an inadequate one, the residual tail falls on the depositor, who should
then be compensated ex-ante with a \emph{risk premium}: a supply yield
above the risk-free rate, the market-discipline mechanism by which
uninsured creditors price institutional risk
\citep{Flannery1998,Egan2017}. I measure it as the trailing-30-day
TVL-weighted stablecoin supply annual percentage yield (APY),
minus the 3-month US Treasury bill rate of $3.70\%$\footnote{FRED series
\texttt{DTB3}, for cross-jurisdiction comparability with the
banking operational-risk literature \citep{Moscadelli2004,Fontnouvelle2006}.
Pools that report a $0\%$ supply APY are excluded. Supply APY includes both
organic lending yield and incentives.}. This
spread is only a proxy for the operational-risk premium, not an
identified price, since the supply yield also reflects utilization, 
rate governance, protocol-idiosyncratic risk factors, and other non-operational factors.

Table~\ref{tab:yield-lending} shows the premium is there in direction
but not in size. Across the unbuffered venues it averages $+54$~basis
points (bps), which on average just covers their own pure premium
($55$~bps mean $\E[S]$, mean coverage $\approx 100\%$) but is a negligible
fraction of the tail-risk premium the per-protocol $\VaR_{99.9}$
implies ($34$--$81\%$ of TVL). That average rests on six venues and hides
wide dispersion: the premium runs from $+257$~bps (Fluid) to $-79$~bps
(Kamino), coverage of the pure premium from $+497\%$ to $-143\%$ with a
median of $60\%$, and three venues yield below the risk-free rate---a
depositor can earn less than the T-bill while bearing an uncapped
operational-risk tail.

\begin{table}[tb]
\centering
\caption{Risk premium (30-day supply APY minus the $3.70\%$ T-bill) for
the ten largest Lending venues by buffer status. The $\E[S]$ and
$\VaR_{99.9}$ columns show pure premium and tail-risk premium; brackets
give the risk premium's coverage of each.}
\label{tab:yield-lending}
\small
\begin{tabular}{l r r r r}
\toprule
Protocol & Total yield & Risk premium & $\E[S]$ & $\VaR_{99.9}$ \\
\midrule
\multicolumn{5}{l}{\emph{No buffer}} \\
Morpho Blue   & 4.60\,\% & $+90$~bps  & $62$~bps ($144\%$)  & $3{,}905$~bps ($+2\%$) \\
JustLend V1   & 3.42\,\% & $-28$~bps  & $60$~bps ($-47\%$)  & $4{,}785$~bps ($-1\%$) \\
Kamino Lend   & 2.91\,\% & $-79$~bps  & $55$~bps ($-143\%$) & $6{,}169$~bps ($-1\%$) \\
Jupiter Lend  & 4.64\,\% & $+94$~bps  & $55$~bps ($172\%$)  & $6{,}605$~bps ($+1\%$) \\
Fluid Lending & 6.27\,\% & $+257$~bps & $52$~bps ($497\%$)  & $6{,}507$~bps ($+4\%$) \\
Euler V2      & 3.59\,\% & $-11$~bps  & $47$~bps ($-23\%$)  & $8{,}131$~bps ($0\%$) \\
\midrule
\multicolumn{5}{l}{\emph{With buffer}} \\
Aave V3         & 3.36\,\% & $-34$~bps  & $64$~bps ($-53\%$)  & $3{,}378$~bps ($-1\%$) \\
SparkLend       & 2.84\,\% & $-86$~bps  & $60$~bps ($-143\%$) & $4{,}634$~bps ($-2\%$) \\
Compound V3     & 2.85\,\% & $-85$~bps  & $55$~bps ($-155\%$) & $6{,}101$~bps ($-1\%$) \\
Venus Core Pool & 2.09\,\% & $-161$~bps & $55$~bps ($-295\%$) & $6{,}226$~bps ($-3\%$) \\
\bottomrule
\end{tabular}
\end{table}

\emph{Both margins together.} Buffer coverage and risk premium are the
two margins at which the same tail can be funded. Because the two
substitute, an efficient market would trade one for the other, buffered
venues paying lower yield. The data run in that direction. Venues with
a buffer pay a significantly lower premium than those without, a median
$-86$~bps against $+39$~bps. The direction is right and statistically
significant (one-sided Mann--Whitney $p=0.01$), but the scale is not.
The whole premium cross-section spans just $418$~bps, while the tail
demands capital of about $18\%$ of TVL, so both margins together leave
it overwhelmingly unfunded, neither within an order of magnitude of
pricing it. The directional gap survives extending the sample to all
Lending venues with net TVL $\ge$~USD~100~m ($n=19$, one-sided
$p=0.03$), and
attenuates below, where posted yields are dominated by airdrop and
other incentive programs rather than by market-cleared compensation for
risk-bearing: the long-tail depositor is chasing incentives, not
pricing risk.

\section{Discussion}\label{sec:discussion}

\emph{An earn product without the capital behind it.} DeFi earn products
are backed by neither a buffer nor a risk premium sized to the tail.
Although the four DeFi core sector severity tails are no heavier than banking's, the
implied capital is larger: e.g. for Lending the per-sector $\VaR_{99.9}$ is
$\sim 18\%$ of TVL, far above the operational-risk
capital banks hold \citep{Fontnouvelle2006}, because the $99.9\%$ year is dominated by the
largest venue's at-risk funds failing. Bridge and Derivatives are larger
still ($32\%$ and $23\%$) but cap-determined, as their $\hat\xi>1$ severity is
unquantifiable without the cap. Of the ten largest Lending venues, four
hold a buffer, covering on average $\sim 5\%$ of their modeled capital
buffer. The six without a buffer carry a supply-yield premium averaging
$+54$~bps, which on average matches their $55$-bps pure premium, though
across only six venues coverage of it ranges from $-143\%$ to $+497\%$. I find evidence that the two responses substitute: buffered venues pay a significantly lower
premium---the trade-off an efficient market would produce---but both
margins are an order of magnitude too small.

\emph{Who bears the shortfall.} The premium I measure is the posted
supply APY, the rate a retail depositor sees and takes. Depositors are
not homogeneous: Cornelli et al.~\cite{Cornelli2025} find that
search-for-yield drives supply, retail deposits in particular, and that
behavior differs by account size. A professional participant can assess
the tail, negotiate bilateral terms,\footnote{Professional market makers
strike bespoke arrangements with protocols through governance, as in
Wintermute's market-making agreement with MakerDAO
(\url{https://thedefiant.io/news/defi/wintermute-makerdao}).} and size
or exit the position. A retail depositor sees only the posted rate and,
lacking standardized disclosure, bears the unpriced tail
disproportionately.

\emph{Policy.}\label{sec:policy} DeFi protocols are not regulated as
banks are, and mandating bank-style operational-risk capital is harder
to implement in a world of decentralized, autonomous DeFi protocols.
What market discipline does require is information: depositors bearing
the operational-risk tail should receive adequate disclosure to judge
whether the yield compensates them. The LDA here supplies the content
of such a disclosure, and I recommend three measures. First,
\emph{standardized loss-and-yield disclosure}, a Pillar~3 analogue that
consolidates the private loss feeds (DefiLlama / SlowMist / de.fi) into
a comparable per-sector loss history reported alongside the supply
yield and its implied risk premium versus the pure premium. Second,
\emph{buffer-and-tail disclosure}: each venue's buffer, the share of
its $\VaR_{99.9}$ allocation covered, and the residual tail the
depositor bears. Third, \emph{periodic recalibration}, because the tail
is non-stationary
\citep{BCBS2017,Flannery1998,Egan2017,Wheatley2016}. The disclosure
duty need not fall on the protocol alone: a decentralized protocol may
have no responsible entity, so the obligation can attach to any service
provider that gives users access (front-ends, wallets, aggregators,
custodial ``earn'' products), the identifiable points at which a retail
depositor actually enters the position.

\emph{A hybrid calibration standard.} Basel's operational-risk framework
runs a two-tier model: standardized business-line coefficients for banks
lacking internal data, and an internal-model option for those with
enough loss history \citep{BCBS2017}. The same hybrid fits DeFi: the
per-sector LDA of Section~\ref{sec:method} is the default, and any
venue with dense enough own-loss history can
substitute a protocol-specific model, disclosed and periodically
recertified.

\emph{Where the capital sits.} Existing on-chain buffers (Aave's
Umbrella, Sky surplus buffer, Compound's reserve factor) all sit inside
the same smart-contract system whose failure they insure, inheriting
the same code-risk. Two decoupling patterns are possible:
\emph{protocol-maintained recovery pools} for the idiosyncratic
component, and \emph{cross-protocol recovery funds}---the on-chain
analogue to post-2008 bank resolution funds---for the systematic
component. Neither is standard practice today.

\emph{Limitations.} Six limitations bound the results.
\emph{Incomplete data.} The analysis rests on publicly reported losses;
sub-threshold and privately resolved incidents are missing, biasing the
body of the severity distribution downward but leaving the large tail
events that drive the capital figures intact.
\emph{Homogeneity.} With no protocol generating enough events for its
own fit, I estimate one severity per sector and allocate it by TVL
share, assuming a common loss process per unit of TVL. A larger or more
complex venue may bear more risk per dollar, which a size-covariate or
hierarchical model would refine. \emph{Recovery.} The figures are gross
of funds later returned; netting recoveries lowers the aggregate, so
the capital estimates are conservative.
\emph{Confounding.} The lower premia at buffered venues are a
descriptive association, not an identified buffer effect. Buffered
venues are also the largest and oldest, so brand and perceived safety
may drive the yield gap as much as the buffer. Perceived safety is
itself an imperfect proxy: industry-standard security audits target
code-level vulnerabilities and offer no protection against the IF and
EDPM categories that account for $26\%$ of Lending losses. \emph{Premium identification.} The supply-yield spread
is a reduced-form proxy, not an identified risk price; it also reflects
utilization, protocol-idiosyncratic risk factors, and rate governance.
A controlled panel regression is the natural extension.
\emph{Stationarity.} The 2020--2026 window is fit as stationary, but
frequency has fallen from the early rugpull wave and severity has
moved toward larger, more concentrated targets; a regime-split fit
would sharpen the point-in-time estimate.

\section{Conclusion}\label{sec:conclusion}

DeFi protocols expose depositors to operational risk
(USD~$9.45$~B across $1{,}075$ events since 2020) but, unlike banks,
need to hold no capital against it, and absent a voluntary buffer the
residual is left with the depositor. Calibrating a Basel LDA gives a
common benchmark, and against it neither margin is sized to the tail:
at the ten largest Lending venues the four disclosed buffers cover on
average $5\%$ of the modeled $\VaR_{99.9}$, and unbuffered venues carry
a supply-yield premium averaging $+54$~bps, short of the tail by more
than an order of magnitude. The two move together in the right direction---the
market prices the risk, but at a magnitude that leaves the tail
overwhelmingly unfunded. Because these protocols are not bank-regulated,
the remedy I draw is not capital mandates but disclosure: standardized
loss, yield, and buffer-coverage information, required of the protocol
or any service provider that fronts access to it.

\begin{credits}
\subsubsection{\discintname}
The author has no competing interests to declare.
\end{credits}

\bibliographystyle{splncs04}
\bibliography{references}

\begin{thebibliography}{10}
\providecommand{\url}[1]{\texttt{#1}}
\providecommand{\urlprefix}{URL }
\providecommand{\doi}[1]{https://doi.org/#1}

\bibitem{Aramonte2021}
Aramonte, S., Huang, W., Schrimpf, A.: {DeFi} risks and the decentralisation
  illusion. BIS Quarterly Review pp. 21--36 (Dec 2021)

\bibitem{Arora2026}
Arora, S., Li, Y., Feng, Y., Xu, J.: A risk scoring framework for critical
  infrastructure of decentralized finance ({DeFi}). Blockchain: Research and
  Applications  (2026). \doi{10.1016/j.bcra.2026.100496}, article 100496

\bibitem{Auer2024}
Auer, R., Haslhofer, B., Kitzler, S., Saggese, P., Victor, F.: The technology
  of decentralized finance ({DeFi}). Digital Finance  \textbf{6}(1),  55--95
  (2024)

\bibitem{BCBS2006}
{Basel Committee on Banking Supervision}: International convergence of capital
  measurement and capital standards: A revised framework, comprehensive version
  (annex 9: Detailed loss event type classification). Tech. rep., Bank for
  International Settlements (2006)

\bibitem{BCBS2017}
{Basel Committee on Banking Supervision}: {Basel III}: Finalising post-crisis
  reforms. Tech. rep., Bank for International Settlements (2017)

\bibitem{BlockSec2026}
{BlockSec}: Weekly {Web3} security incident roundups.
  \url{https://blocksec.com/blog} (2026), 15 weekly posts ingested, accessed
  2026-05-18

\bibitem{BockerKluppelberg2005}
B\"ocker, K., Kl\"uppelberg, C.: Operational {VaR}: a closed-form
  approximation. Risk  \textbf{18}(12),  90--93 (2005)

\bibitem{BuhlmannGisler2005}
B\"uhlmann, H., Gisler, A.: A Course in Credibility Theory and Its
  Applications. Universitext, Springer, Berlin (2005)

\bibitem{Chernobai2008}
Chernobai, A.S., Rachev, S.T., Fabozzi, F.J.: Operational Risk: A Guide to
  {Basel II} Capital Requirements, Models, and Analysis. Wiley (2007)

\bibitem{Cope2009}
Cope, E.W., Mignola, G., Antonini, G., Ugoccioni, R.: Challenges and pitfalls
  in measuring operational risk from loss data. Journal of Operational Risk
  \textbf{4}(4),  3--27 (2009)

\bibitem{Cornelli2025}
Cornelli, G., Gambacorta, L., Garratt, R., Reghezza, A.: Why {DeFi} lending?
  evidence from {Aave V2}. Journal of Financial Intermediation  \textbf{63}
  (2025). \doi{10.1016/j.jfi.2025.101166}, article 101166

\bibitem{DeFiRektDb2026}
{de.fi (formerly DeFiYield)}: rekt-database. REST API
  \url{https://api.de.fi/v1/rekt/list}; UI \url{https://de.fi/rekt-database}
  (2026), retrieved 2026-05-29

\bibitem{Defillama}
{DefiLlama}: Public hack ledger and protocol catalog.
  \url{https://api.llama.fi/hacks} (2026), accessed 2026-05-16

\bibitem{DuMouchel1983}
DuMouchel, W.H.: Estimating the stable index $\alpha$ in order to measure tail
  thickness: A critique. Annals of Statistics  \textbf{11}(4),  1019--1031
  (1983)

\bibitem{Egan2017}
Egan, M., Horta\c{c}su, A., Matvos, G.: Deposit competition and financial
  fragility: Evidence from the {U.S.} banking sector. American Economic Review
  \textbf{107}(1),  169--216 (2017)

\bibitem{Eling2019}
Eling, M., Wirfs, J.: What are the actual costs of cyber risk events? European
  Journal of Operational Research  \textbf{272}(3),  1109--1119 (2019)

\bibitem{EmbrechtsKM1997}
Embrechts, P., Kl\"uppelberg, C., Mikosch, T.: Modelling Extremal Events for
  Insurance and Finance. Springer (1997)

\bibitem{Flannery1998}
Flannery, M.J.: Using market information in prudential bank supervision: A
  review of the {U.S.} empirical evidence. Journal of Money, Credit and Banking
   \textbf{30}(3),  273--305 (1998)

\bibitem{Fontnouvelle2006}
de~Fontnouvelle, P., Rosengren, E., Jordan, J.: Implications of alternative
  operational risk modeling techniques. In: Carey, M., Stulz, R.M. (eds.) The
  Risks of Financial Institutions, pp. 475--505. University of Chicago Press
  (2007)

\bibitem{Heimbach2022}
Heimbach, L., Schertenleib, E., Wattenhofer, R.: Risks and returns of {Uniswap
  V3} liquidity providers. In: Proceedings of the 4th ACM Conference on
  Advances in Financial Technologies (AFT). pp. 89--101 (2022)

\bibitem{Kismp2026}
{Kismp123 et al.}: {DeFi-Security-Incident}: A comprehensive reference of
  real-world {DeFi} security incidents with root-cause analysis, attack flow,
  and {PoC} code. GitHub repository,
  \url{https://github.com/kismp123/DeFi-Security-Incident} (2026), 2020--2026,
  accessed 2026-05-18

\bibitem{Moscadelli2004}
Moscadelli, M.: The modelling of operational risk: Experience with the analysis
  of the data collected by the {Basel} committee. Temi di Discussione (Working
  Paper)~517, Banca d'Italia (2004)

\bibitem{Perez2021}
Perez, D., Livshits, B.: Smart contract vulnerabilities: Vulnerable does not
  imply exploited. In: USENIX Security Symposium (2021)

\bibitem{Qin2021}
Qin, K., Zhou, L., Livshits, B., Gervais, A.: Attacking the {DeFi} ecosystem
  with flash loans for fun and profit. In: Financial Cryptography and Data
  Security (FC) (2021)

\bibitem{Rekt2026}
{Rekt News}: Rekt leaderboard. \url{https://rekt.news/leaderboard} (2026),
  2020--2026, accessed 2026-05-18

\bibitem{SlowMistHacked2026}
{SlowMist Team}: {SlowMist} hacked. Blockchain-event tracker,
  \url{https://hacked.slowmist.io} (2026), retrieved 2026-05-29

\bibitem{DeFiHackLabs2026}
{SunWeb3Sec et al.}: {DeFiHackLabs}: Reproduce {DeFi} hacked incidents using
  {Foundry}. GitHub repository,
  \url{https://github.com/SunWeb3Sec/DeFiHackLabs} (2026), 2021--2026, accessed
  2026-05-18

\bibitem{Vuong1989}
Vuong, Q.H.: Likelihood ratio tests for model selection and non-nested
  hypotheses. Econometrica  \textbf{57}(2),  307--333 (1989)

\bibitem{Weingartner2023}
Weing\"artner, T., Fasser, F., Reis S\'a~da Costa, P., Farkas, W.: Deciphering
  {DeFi}: A comprehensive analysis and visualization of risks in decentralized
  finance. Journal of Risk and Financial Management  \textbf{16}(10), ~454
  (2023). \doi{10.3390/jrfm16100454}

\bibitem{Werner2022}
Werner, S.M., Perez, D., Gudgeon, L., Klages-Mundt, A., Harz, D., Knottenbelt,
  W.J.: {SoK}: Decentralized finance ({DeFi}). In: Proceedings of the 4th ACM
  Conference on Advances in Financial Technologies (AFT) (2022)

\bibitem{Wheatley2016}
Wheatley, S., Maillart, T., Sornette, D.: The extreme risk of personal data
  breaches and the erosion of privacy. European Physical Journal B
  \textbf{89}(1), ~7 (2016)

\bibitem{Zhou2023}
Zhou, L., et~al.: {SoK}: Decentralized finance ({DeFi}) attacks. In: IEEE
  Symposium on Security and Privacy (S\&P) (2023)

\end{thebibliography}

\end{document}